\documentclass{article}

\usepackage{arxiv}
\usepackage[sort&compress,numbers]{natbib}

\usepackage{soul}

\usepackage[utf8]{inputenc} 
\usepackage[T1]{fontenc}    
\usepackage{hyperref}       
\usepackage{url}            
\usepackage{booktabs}       
\usepackage{amsfonts}       
\usepackage{nicefrac}       
\usepackage{microtype}      
\usepackage{lipsum}		
\usepackage{graphicx}
\usepackage{doi}
\usepackage{amssymb}
\usepackage{xcolor}
\usepackage{comment}
\usepackage{amsmath}

\DeclareMathOperator*{\argmin}{arg\,min}
\usepackage{placeins}
\usepackage[capitalise]{cleveref}

\title{Transcending shift-invariance in the paraxial regime via end-to-end inverse design of\\ freeform nanophotonics}

\author{ {\bf{\hspace{1mm}William F.~Li\textsuperscript{1,*}, Gaurav Arya\textsuperscript{2}, Charles Roques-Carmes\textsuperscript{1}, Zin Lin\textsuperscript{3}, Steven G.~Johnson\textsuperscript{2}, Marin Solja\v{c}i\'{c}\textsuperscript{1,4}}}\\
\textsuperscript{1}\normalfont{ Research Laboratory of Electronics, Massachusetts Institute of Technology, Cambridge, MA 02139}\\
\textsuperscript{2}\normalfont{ Department of Mathematics, Massachusetts Institute of Technology, Cambridge, MA 02139}\\
\textsuperscript{3}\normalfont{ Bradley Department of Electrical and Computer Engineering, Virginia Tech, Blacksburg, VA 24060}\\
\textsuperscript{4}\normalfont{ Department of Physics, Massachusetts Institute of Technology, Cambridge, MA 02139}\\
\textsuperscript{*}wfli@mit.edu
}

\renewcommand{\headeright}{ }
\renewcommand{\undertitle}{ }

\hypersetup{
pdftitle={Transcending Shift-Invariance Freeform Nanophotonics},
pdfsubject={photonics},
pdfauthor={W.F. Li, G. Arya, C. Roques-Carmes, Z. Lin, S.G. Johnson, M. Soljacic},
pdfkeywords={Paraxial optics, topology optimization},
}

\begin{document}
 \setlength{\headheight}{22.37994pt}
 \addtolength{\topmargin}{-8.37994pt}
\maketitle

\begin{abstract}
Traditional optical elements and conventional metasurfaces obey shift-invariance in the paraxial regime. For imaging systems obeying paraxial shift-invariance, a small shift in input angle causes a corresponding shift in the sensor image. Shift-invariance has deep implications for the design and functionality of optical devices, such as the necessity of free space between components (as in compound objectives made of several curved surfaces). We present a method for nanophotonic inverse design of compact imaging systems whose resolution is not constrained by paraxial shift-invariance. Our method is end-to-end, in that it integrates density-based full-Maxwell topology optimization with a fully iterative elastic-net reconstruction algorithm. By the design of nanophotonic structures that scatter light in a non-shift-invariant manner, our optimized nanophotonic imaging system overcomes the limitations of paraxial shift-invariance, achieving accurate, noise-robust image reconstruction beyond shift-invariant resolution.

\end{abstract}


\section{Introduction}

\begin{figure}
    \begin{center}
    \includegraphics[width=\textwidth]{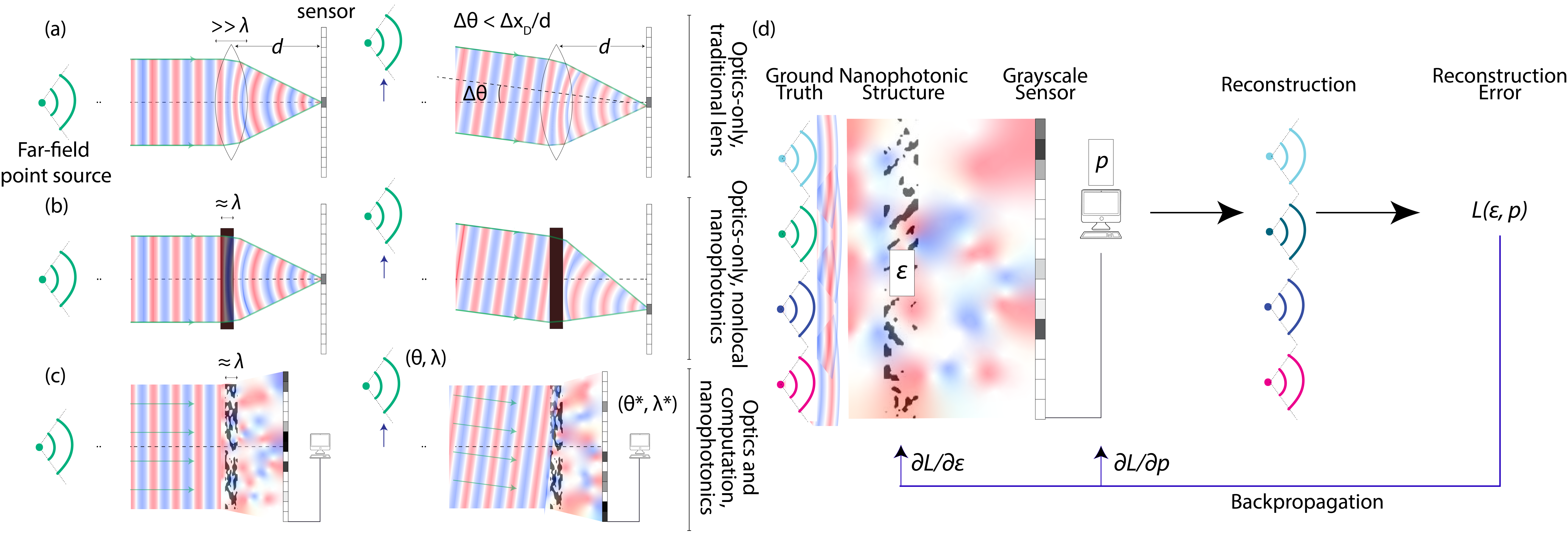}
    \caption{\textbf{Transcending shift-invariance with end-to-end optimized freeform metasurfaces.} (a) Left (resp. Right): Image formation with a conventional thin lens at normal (resp. oblique) incidence. The two angles are not resolved. (b) Left (resp. Right): Image formation with a designed nonlocal nanophotonic optical element at normal (resp. oblique) incidence. The two angles are resolved. (c) Left (resp. Right): Image formation in an end-to-end nanophotonic pipeline at normal (resp. oblique) incidence. The two angles are resolved with computational processing. (d) The end-to-end design pipeline presented in this work. From left to right: the ground truth emits polychromatic light scattered through a nanophotonic structure and measured on a grayscale sensor. The signal on the sensor is then fed to a reconstruction algorithm with hyperparameters $p$, which outputs a reconstruction of the original signal. During training, a reconstruction error $L(\varepsilon, p)$ is computed as a function of the nanophotonic structure permittivity profile $\varepsilon$ and the reconstruction hyperparameters $p$. To design the structure $\varepsilon$, gradients $\partial L/\partial \varepsilon $ are backpropagated back to the nanophotonic structure. To tune the hyperparameters $p$, gradients $\partial L/\partial p $ are propagated back to the reconstruction algorithm.}
    \label{fig:1}
    \end{center}
\end{figure}

The design of imaging systems that transcend paraxial shift-invariance is the next step in making compact, high-resolution imagers. Conventional optical elements, such as thin lenses \cite{goodman2005introduction}, obey the property of paraxial shift-invariance, meaning the best angular resolution they can achieve is $\Delta \theta \sim \Delta x_D/d$, where $d$ is the distance between the optical element and the sensor and $\Delta x_D$ is the width of a detector pixel on the sensor (Fig \ref{fig:1}a, left and right). Traditional metasurfaces relying on pre-computed paraxial phase libraries \cite{guo2020squeeze, yu2014flat, arbabi2015dielectric, khorasaninejad2016metalenses, chen2016review, engelberg2020advantages} are also constrained by the same limitation. 
Here, we present a method for the design of freeform nanophotonic optical elements that overcome such constraints on angular resolution. We demonstrate our method to design two-dimensional (2D) and three-dimensional (3D) freeform nanophotonic structures for angle-resolved spectrometry at angular resolutions beyond what is allowed by the paraxial limit. We design our freeform nanophotonic structures through topology optimization \cite{jensen2011topology, christiansen2021inverse, molesky2018inverse} in an end-to-end \cite{arya2022end, sitzmann2018end, lin2021end, lin2022end, sun2020end, tseng2021differentiable} pipeline (Fig.~\ref{fig:1}(d)), which directly minimizes the ultimate reconstruction error. In our approach, freeform nanophotonic geometries are parametrized by dielectric permittivity $\varepsilon$ at every pixel in a design region, amounting to tens of thousands of optimization parameters. We show that our optimized structures outperform both a conventional thin lens (which obeys paraxial shift-invariance) and random nanophotonic structures (which are not beholden to the same limit).

 Previously, the challenge of transcending paraxial shift-invariance has been addressed with super-cell metasurfaces \cite{spagele2021multifunctional} and nonlocal optics \cite{guo2020squeeze, reshef2021optic, overvig2022diffractive}, in the context of ``space squeezing"  -- compression of free space by designed nanophotonic structures. For a given detector pixel size $\Delta x_D$ in traditional optical elements, the paraxial shift-invariance limit dictates the distance $d$ between the optical element and sensor required to capture images at a given spatial resolution $\Delta \theta$, where higher resolutions with $\Delta \theta < \Delta x_D/d$ cannot be captured by an imaging system with a conventional thin lens (\cref{fig:1}(a)). Minimizing free space in imaging systems, or ``space squeezing," is part of a broader effort to reduce the volume of imaging systems. Prior work seeks to minimize the volume of both the optical element \cite{yu2014flat, arbabi2015dielectric, khorasaninejad2016metalenses, chen2016review} and free-space \cite{guo2020squeeze, reshef2021optic} separately, and typically involves two engineered structures: a local metasurface to replace the lens and a nonlocal structure to replace free space (e.g. a multi-layer stack acting as a space squeezer). This two-structure system is necessary because traditional metasurfaces are characterized by local transfer functions while replacing free space requires a nonlocal (momentum-dependent) transfer function {\cite{guo2020squeeze}}. Nonlocal optimized metasurfaces can resolve angles within the paraxial regime, as shown in Fig.~\ref{fig:1}(b). Our method, shown in Fig.~\ref{fig:1}(c), offers a more compact alternative to space squeezing by designing a single thin engineered nanophotonic structure which, in conjunction with a computational-imaging algorithm, replaces both the lens and free space, rather than one for each. Unlike prior approaches that separate the two problems (lens and free space), our approach combines them into a single end-to-end image-reconstruction problem.

In addition to the nanophotonic structure design, a robust computational reconstruction component is essential to imaging beyond the paraxial limit. Prior work realized compact imagers in inverse-designed optics-only systems~\cite{lin2021computational}. Such techniques rely on the optimization of many degrees of freedom (typically distributed over an entire optimization volume) to realize a pre-defined optical functionality. In contrast, our work leverages recent developments in end-to-end inverse-design: harnessing computational reconstruction to loosen constraints on the transformation imparted by the optimized optical elements. We achieve this by resolving images and patterns that would be unreadable to the human eye (and could not be pre-defined by the user as an optimization task). This allows for the design of thinner, less complex nanophotonic structures that need only to produce an image interpretable by the reconstruction algorithm. In previous work in end-to-end optimization, nanophotonic structures have been paired with various image reconstruction algorithms, including neural networks, compressed sensing (or Lasso-regularized regression), Tikhonov-regularized regression, and elastic-net regression \cite{Yanny2020, 7882664, 10.5555/3104322.3104374, istanet, markley2021physicsbased}. In our work, we use the elastic-net reconstruction algorithm, which combines the Lasso ($l_1$) regularization term and the Tikhonov ($l_2$) regularization term. Intuitively, the $l_1$ term encourages the regression to deliver a sparse solution, which is of particular interest to us in the detection of angle and frequency of incoming laser beams under the often reasonable assumption of there coming only a few beams at a time. We give the reconstruction algorithm flexibility to choose whether to emphasize the Lasso or Tikhonov regularization terms by optimizing the elastic-net hyperparameters \cite{Bertrand_Klopfenstein_Blondel_Vaiter_Gramfort_Salmon20}; we show that a sparse problem generally results in the optimization greatly emphasizing the Lasso term and de-emphasizing the Tikhonov term. Our reconstruction algorithm is paired with the topology-optimized structure, allowing for the automated discovery of both freeform designs and reconstruction hyperparameters.

\section{End-to-end Optimization Pipeline}
\label{sec:methods}

\subsection{Image formation}
We consider polychromatic and spatially-extended objects, therefore describing the ground-truth as a multi-dimensional tensor with at most 4 dimensions (3D space + 1D spectral). For convenience, we represent this tensor as a flattened single vector $\mathbf{u}$ where each component corresponds to a unique angle-frequency pair $(\theta, \lambda)$. We propagate the ground truth object through our nanophotonic structure and free space to generate a raw, noisy, grayscale image at the detector $\mathbf{v}$, where 
\begin{equation}
\mathbf{v} = G\left(\varepsilon(\mathbf{r})\right)\mathbf{u} + \eta. 
\end{equation}
In the above expression, $\varepsilon(\mathbf{r})$ is the dielectric profile of the nanophotonic structure, $G\left(\varepsilon(\mathbf{r})\right)$ is the measurement matrix of the imaging system (a function of $\varepsilon(\mathbf{r})$), and $\eta$ is the additive Gaussian noise, whose standard deviation is proportional to the average intensity on the detector.
The degrees of freedom of the structure $\varepsilon(\mathbf{r})$ are free to take on arbitrary designs through the optimization, and, in particular, we do not assume that $G$ should result in shift-invariant point-spread function (PSF). This is key for allowing our system to differentiate between angles in the paraxial regime. We emphasize that such differentiation is not possible under the assumption of shift-invariance, which is commonly used as a computational simplification \cite{goodman2005introduction}, as is shown in Fig.~\ref{fig:1}(a,b) We numerically compute the measurement matrix $G$ from $\varepsilon(\mathbf{r})$ using the finite-difference time-domain (FDTD) method \cite{oskooi2010meep}.

\subsection{Parameter estimation}
Our pipeline is made up of a nanophotonic structure and a reconstruction algorithm. First, the electric fields generated on the detector by the ground truth object $\mathbf{u}$ are calculated with the image formation process, as described in the previous section. The result of that first step is the raw, noisy vector of intensities $\mathbf{v}$. We feed this vector into the computational reconstruction algorithm. The computational reconstruction algorithm uses elastic-net regression to reconstruct an object $\mathbf{u}_\text{est}.$ Elastic-net regression is a form of linear regression with both a Lasso ($l_1$) and a Tikhonov ($l_2$) regularization term. Mathematically, the reconstruction problem amounts to solving the following optimization problem:
\begin{equation}\mathbf{u}_\text{est} = \argmin_\mathbf{u} \left(\left\Vert\mathbf{v} - G(\varepsilon(\mathbf{r}))\mathbf{u}\right\Vert_2^2 + \lambda_1 \left\Vert \mathbf{u}\right\Vert_2^2 + \lambda_2\left\Vert\mathbf{u}\right\Vert_1 \right).\end{equation} 

$\lambda_1$ and $\lambda_2$ are the reconstruction hyperparameters. $\lambda_2$ controls the magnitude of the Lasso ($l_1$) regularization term, which selects for sparsity. Our end-to-end optimization task is to find values for $\varepsilon(\mathbf{r})$, $\lambda_1$, and $\lambda_2$ that minimize the normalized reconstruction error averaged over the training set and corresponding image noise $$L(\mathbf{u}_\text{est}, \mathbf{u}) = \left\langle \frac{\left\Vert\mathbf{u} - \mathbf{u}_\text{est} \right\Vert_2^2}{\left\Vert\mathbf{u}\right\Vert_2^2}\right\rangle_{\mathbf{u}, \eta}.$$ Our end-to-end optimization task to computationally design an imaging system can therefore be written mathematically as: 

\begin{equation}\varepsilon(\mathbf{r})^{(\text{opt})}, \lambda_1^{(\text{opt})}, \lambda_2^{(\text{opt})} = \argmin_{\varepsilon(\mathbf{r}), \lambda_1, \lambda_2}L(\mathbf{u}, \mathbf{u}_\text{est}).\end{equation}

To perform optimization in the end-to-end framework, we need to compute gradients for the loss $L$ with respect to  the parameters $\varepsilon(\mathbf{r})$, $\lambda_1$, and $\lambda_2$. Gradients are back-propagated through the elastic-net reconstruction algorithm by finding the derivatives of the Karush-Kuhn-Tucker conditions \cite{arya2022end}; the gradients are then back-propagated through the FDTD simulation using an adjoint simulation. We train our end-to-end system in a two-step process: first, we use the method of moving asymptotes (MMA) \cite{sgjnlopt, svanberg2002class} on a fixed training set for topology optimization of the nanophotonic structure $\varepsilon(\mathbf{r})$; then, we use Adam optimization \cite{kingma2014adam} on data randomly generated each iteration to tune the reconstruction hyperparameters $\lambda_1$ and $\lambda_2$. We benchmark our optimized designs against randomly initialized designs and thin lens designs by optimizing $\lambda_1$ and $\lambda_2$ for each design and comparing the optimized loss $L$. Because we randomly generate new data for each iteration of Adam, the training loss serves also as validation loss.

\section{Results}
\label{sec:results}

\begin{figure}
\begin{center}
\includegraphics[width=\textwidth]{"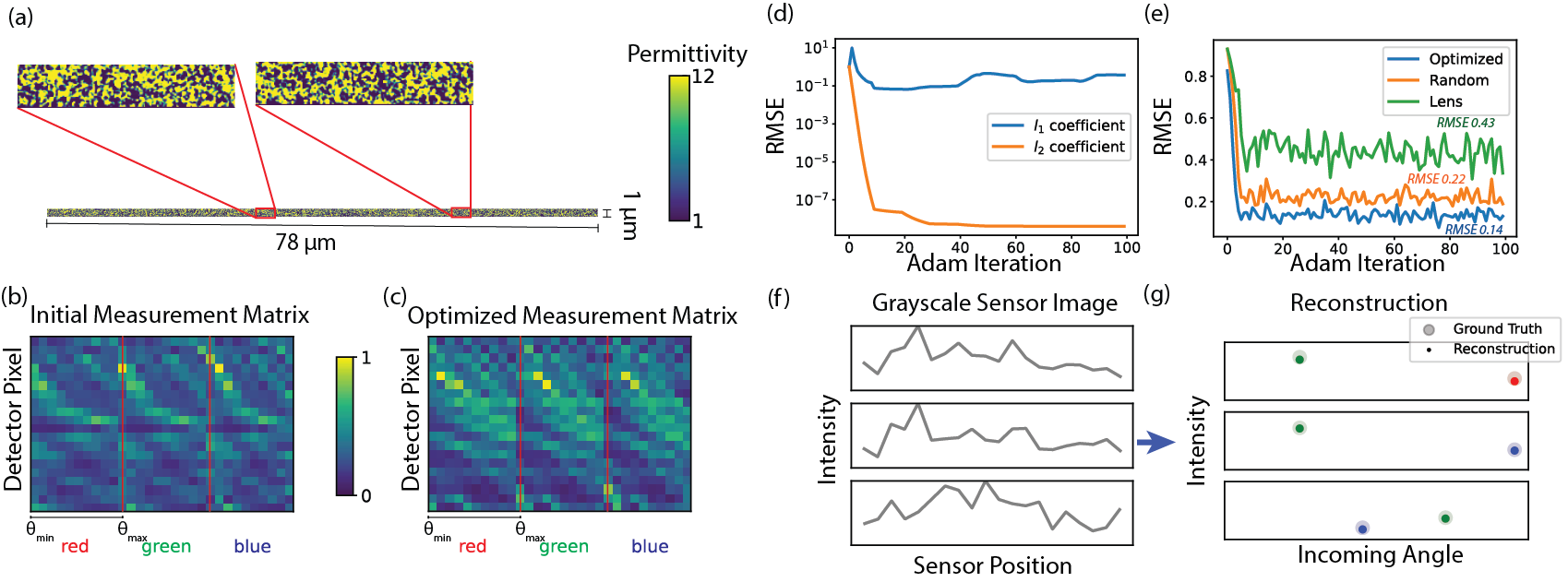"}
\end{center}
\caption{\textbf{Sparse, underdetermined angle-resolved spectrometry.} (a) The optimized $78 \, \mathrm{\mu m }\times 1 \, \mathrm{\mu m }$ volumetric nanophotonic structure, represented as a binary heatmap of $\varepsilon(\mathbf{r})$, with zoom-ins on two regions of the structure. There are 1296 design degrees of freedom per $1 \, \mathrm{\mu m}^2$. In the final optimized structure, $\varepsilon(\mathbf{r})$ takes on permittivities of only $1$ and $12$. (b, c) Measurement matrix $G$ (b) before and (c) after optimization, with detector pixel position on the $y$-axis and incoming angle and color on the $x$-axis. (d) Evolution of the $l_1$ (Lasso) and $l_2$ (Tikhonov) regularization coefficients over optimization. (e) Convergence of the reconstruction error during end-to-end optimization (blue) compared to convergence of the reconstruction error during reconstruction-only optimization with a random structure (orange). (f) Grayscale images formed from the ground truths in \cref{fig:2}(g) on the detector with 1\% noise. (g) Sparse ground truth signal of $2$ activated elements out of a total of $10 \, \text{angle} \times 3 \, \text{frequency}$ possibilities, overlaid with reconstructed sparse signal from the image in \cref{fig:2}(f).}

\label{fig:2}
\end{figure}

\begin{figure}
\begin{center}
\includegraphics[width=\textwidth]{"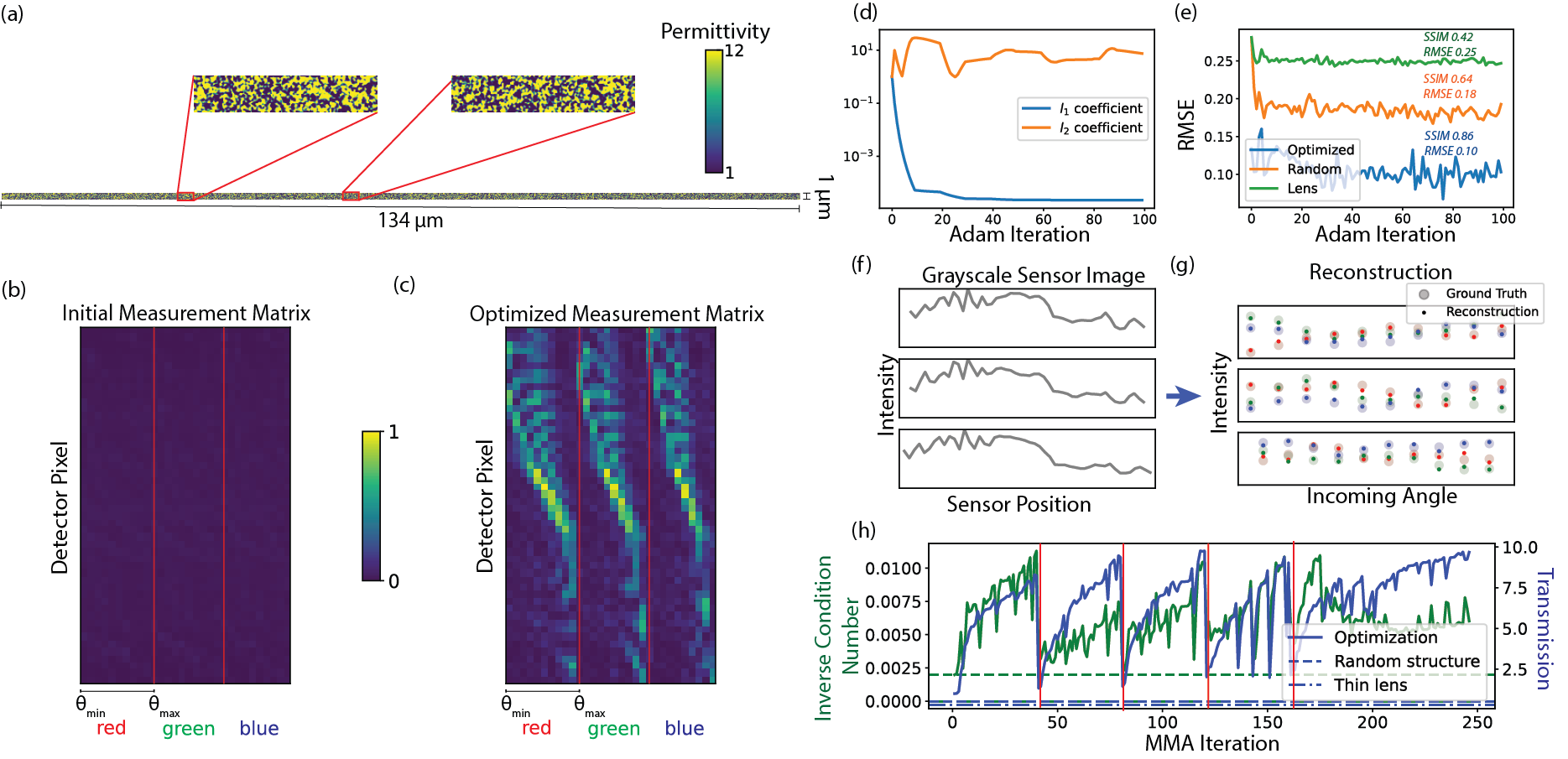"}
\end{center}
\caption{\textbf{General angle-resolved spectrometry on a large sensor.} (a) The optimized $134 \, \mathrm{\mu m }\times 1 \, \mathrm{\mu m }$ volumetric nanophotonic structure, represented as a binary heatmap of $\varepsilon(\mathbf{r})$, with zoom-ins on two regions of the structure. There are 1296 design degrees of freedom per $1 \, \mathrm{\mu m}^2$. In the final optimized structure, $\varepsilon(\mathbf{r})$ takes on binary permittivities of only $1$ and $12$. (b) The measurement matrices $G$ (b) before and (c) after optimization. (d) Evolution of the $l_1$ (Lasso) and $l_2$ (Tikhonov) regularization coefficients over optimization. (e) Convergence of the reconstruction error during end-to-end optimization. (f) Grayscale images formed from the ground truths in \cref{fig:3}(g) on the detector with 1\% noise. (g) Ground truth signal of $10 \, \text{angles} \times 3 \, \text{frequencies}$ overlaid with reconstructed sparse signal from the image in \cref{fig:3}(f). (h) Comparison of inverse condition numbers and transmissions between optimized structure, random structure, and thin lens. Red vertical lines indicate new training phases, increasing binary filter strength on the structure with each new phase. }
\label{fig:3}
\end{figure}

We showcase our end-to-end pipeline in three types of reconstruction problems where, in order to accurately reconstruct the ground truth, we need finer angular resolution than what can be offered by paraxial optics. In all three reconstruction problems, we show angular resolutions that, in a paraxial system, would conventionally require over 20 times the separation we use between the lens and sensor ($d$), observing the measurement matrices breaking shift-invariance to resolve angles in the paraxial regime. In the overdetermined case, we also refer to this effect as ``space squeezing." We define the ``compression ratio" as the factor by which we reduce the separation between the lens and sensor from the minimum required in paraxial imaging. In the first reconstruction problem (Fig.~\ref{fig:2}), we solve the 2D ``sparse sensing" problem: an underdetermined inverse problem in two dimensions with a sparse prior, which means we assume there are far fewer nonzero elements in the ground truth than the total possible size of the ground truth; for instance, we may be detecting the angle and frequency of incoming laser beams, with the assumption that there are a relatively few number of incoming beams at any given time. The flexibility of our elastic-net reconstruction along with the requirement of sparsity allows us to accurately recover the ground truth signal. Over the course of optimization, our reconstruction algorithm settles into the compressed sensing limit of elastic-net, increasing the $l_1$ (Lasso) normalization coefficient and shrinking the $l_2$ (Tikhonov) normalization coefficient. In the second reconstruction problem (Fig.~\ref{fig:3}), we solve the 2D space squeezing problem: an overdetermined inverse problem in two dimensions with no sparsity prior. Here, the image reconstruction algorithm emphasizes the $l_2$ coefficient instead. In the third reconstruction problem (Fig.~\ref{fig:4}), we generalize the space squeezing problem to three dimensions. Throughout the results, we report condition numbers of the measurement matrix $G(\varepsilon(\mathbf{r}))$. Intuitively, a lower condition number means that the matrix is more robust to noise for reconstruction. Qualitatively, we find that the most reconstruction improvement for condition numbers that start in the 100s or 1000s and decrease by a factor of $1.5$ or more. 

We perform the 2D optimizations with 240 CPUs over the span of 2 to 3 days, and we perform the 3D optimizations with 480 CPUs over the span of 3 to 4 days. During topology optimization, we optimize the value of $\epsilon(\mathbf{r})$ continuously at each pixel, gradually turning on a binary filter over the course of optimization. For the 2D problems, we also gradually turn on a Gaussian filter to increase the feature size of the nanophotonic structure.

\subsection{Two-dimensional Sparse Spectral-Angular Sensing}
In this example, we show how end-to-end optimization can be used to reconstruct the spectrum and angle of incidence of an object beyond the paraxial limit for sparse ground-truth objects. Sparse sensing has application to laser awareness, enabling the simultaneous sensing of a small number of distinct signals from different directions and frequencies. The ground-truth object has dimensions $10 \,\text{angles} \times 3\, \text{frequencies}$, making the vector representation of the ground-truth $\mathbf{u}$ a 30-component vector. The angles are uniformly spaced between $-0.04$ radians and $0.04$ radians from normal incidence, such that the nonzero angles for each ground-truth $\mathbf{u}$ are drawn from this particular set of $10$ angles. The frequencies correspond to red ($672 \, \mathrm{nm}$), green ($560 \, \mathrm{nm}$), and blue light ($448 \, \mathrm{nm}$). Each detector pixel has length $ x_D = 3.36 \, \mathrm{\mu m}$. We set the sensor $d=11.2 \, \mathrm{\mu m}$ from the structure. In the sparse sensing problem, we optimize the nanophotonic structure in a $78 \, \mathrm{\mu m} \times 1\, \mathrm{\mu m}$ design region. 
We first demonstrate our method in a 2D setting. Here we use a sensor with $20$ detector pixels, which is $2/3$ times the total number of parameters to be reconstructed. From the specifications above, this makes the entire sensor $67.2 \, \mathrm{\mu m}$ long. We give our nanophotonic structure a design region of size $78 \, \mathrm{\mu m} \times 1 \, \mathrm{\mu m}$ (\cref{fig:2}(a)), with 1296 degrees of freedom per $1 \, \mathrm{\mu m}^2$. 

We initialize the nanophotonic structure as a random binary structure. This results in the measurement matrix shown in Fig.~\ref{fig:2}(b). The measurement matrix is constructed by propagating monochromatic plane waves that span the length of the nanophotonic structure and measuring the result on the sensor for each plane wave. The measurement matrix is then indexed by sensor pixel along the rows and by incoming angle and frequencies along the columns. The vertical red lines separate the matrix into sections by incoming plane wave frequency; within each section, the incoming angles range over all $10$ angles. Because the sensor is smaller than the structure and therefore smaller than the full span of the incoming plane waves, the entire sensor initially detects near-constant low intensity over all incoming angles and frequencies. 

After optimization, using the procedure described in the previous section, the measurement matrix is shown in Fig.~\ref{fig:2}(c) and the optimized structure in \cref{fig:2}(a). Qualitatively, the optimized matrix has less correlation between adjacent pixels than the initial measurement matrix, which matches findings that matrices with independent random pixels are well-suited for compressed sensing problems \cite{candes2008introduction}. The sensitivity (or robustness) of the reconstruction to environmental noise can be characterized using the condition number of the measurement matrix, defined as the ratio of maximal to minimal singular value. Quantitatively, over the course of the optimization, the condition number of the measurement matrix decreases from 2431 to 1514, and the optical transmission increases from 0.21 to 0.27 (Table 1). We note that the columns of the optimized measurement matrix (\cref{fig:2}(c)) differ substantially from each other, meaning the imaging system is not shift-invariant in the angles. This property is key for accurately resolving adjacent angles in the paraxial regime. By comparison, the columns for the measurement matrix of a thin lens are the same across different incoming angles. Meanwhile, on the reconstruction side, the optimization emphasizes the $l_1$ regularization coefficient while making the $l_2$ coefficient shrink to a negligible value, nearly five orders of magnitude under the $l_1$ coefficient (Fig.~\ref{fig:2}(d)). Here, the presence of the $l_2$ regularization term improves convergence in the early iterations of reconstruction hyperparameter optimization even though it is eventually set to nearly zero, as observed in \cite{arya2022end}. Altogether, by giving the end-to-end optimization an underdetermined sparse problem, the optimization computationally settles on both Lasso regression and a nanophotonic structure that leads to a randomized measurement matrix, conditions consistent with existing compressed sensing literature. We emphasize that we specified neither of those conditions as explicit goals of our optimization.

Overall, the optimized system takes in the input signal from a sparse, multichromatic ground truth, forming a noisy, randomized, grayscale image (Fig.~\ref{fig:2}f), and accurately recovering the ground truth by solving the compressed sensing problem with Lasso regression (Fig.~\ref{fig:2}g). The optimized end-to-end system accurately recovers incoming sparse signals under $1\%$ sensor noise with RMSE $0.14$, a significant improvement over a random structure (RMSE $0.22$) and a thin lens focusing to the detector (RMSE $0.43$) (Fig.~\ref{fig:2}e).

From the parameters described previously, the interval between adjacent angles is $\Delta \theta = \frac{\pi}{360}$. This puts us well beyond the paraxial limit, as $\tan\left(\Delta \theta\right) \approx 0.0087 < 0.30 = \frac{x_D}{d}$. In terms of space squeezing, this gives us an effective compression ratio $\frac{x_D/d}{\tan\left(\Delta \theta\right)}$ of $34.5$ over an angular bandwidth of 0.08 radians and a spectral bandwidth of $2.2\cdot 10^{14}$ Hz. For a comparison of angular and spectral ranges, see Table 2. Because we are beyond the paraxial limit, this is a situation where a traditional lens would do poorly---in particular, multiple adjacent angles often give the same reading on the sensor with a traditional lens, illustrated in Fig. \ref{fig:1}.

\subsection{Two-dimensional Polychromatic Space Squeezing}
\label{2D_squeeze}
This second reconstruction problem shares many of the same design parameters as the other two-dimensional problem described in section 3.1 (same ground-truth vector dimensions, angle range, frequencies, and detector parameters). However, in the case of general-purpose space squeezing, one cannot \textit{a priori} assume sparsity of the ground truth image. 

We therefore demonstrate our method in an overdetermined system without the sparse prior. Here we activate all 30 components of the 10 angles $\times$ 3 frequencies ground-truth object. We use a sensor with 50 pixels, so our reconstruction algorithm is solving an overdetermined regression problem. From the specifications, the overall sensor is then $168 \, \mathrm{\mu m}$ long. Our nanophotonic design region is extended to $134  \, \mathrm{\mu m} \times 1 \, \mathrm{\mu m}$ to better match the sensor size (\cref{fig:3}(a)), again with 1296 degrees of freedom per $1 \, \mathrm{\mu m}^2$. . We again initialize the nanophotonic structure as a random structure, leading to the measurement matrix \cref{fig:3}(b) and the structure depicted in  \cref{fig:3}(a). After optimization, the measurement matrix increases significantly in maximum intensity and is no longer shift-invariant (\cref{fig:3}(b,c)). Quantitatively, over the optimization, the condition number decreases from 499 to 166, and the transmission increases from 0.20 to 0.53 (Table 1, \cref{fig:3}(h)). Here sparsity is not enforced in the ground-truth, so the regularization coefficients no longer emphasize the $l_1$ term in the hyperparameter tuning (\cref{fig:3}(d)). An example ground truth, noisy grayscale image, and reconstructed signal are shown in \cref{fig:3}(f,g). The optimized system has structural similarity index measure, or SSIM, 0.86 and RMSE 0.10 at 1\% Gaussian image noise, a significant improvement over a random structure (SSIM 0.64, RMSE 0.18) and a lens (SSIM 0.42, RMSE 0.25). 

To go from the sparse underdetermined problem to the general overdetermined problem, we only had to change the physical conditions of the pipeline. In particular, we made no changes to the initial reconstruction hyperparameters, with the optimization automatically choosing to emphasize the $l_1$ regularization term and reduce the $l_2$ term to nearly zero, the opposite of what we had previously observed in \cref{fig:2}(d). The flexibility of elastic-net and end-to-end optimization allows us to solve these different classes of problems without any manual tuning of the reconstruction algorithm. We also emphasize the application of the general overdetermined problem for space squeezing, or imaging in systems with compact free space. 

Here, $\frac{x_D}{d} = 0.30$ and the compression ratio is 34.5, same as in the previous example.

\subsection{Three-dimensional Polychromatic Space Squeezing}
\begin{figure}
\centering
\includegraphics[width=\textwidth]{"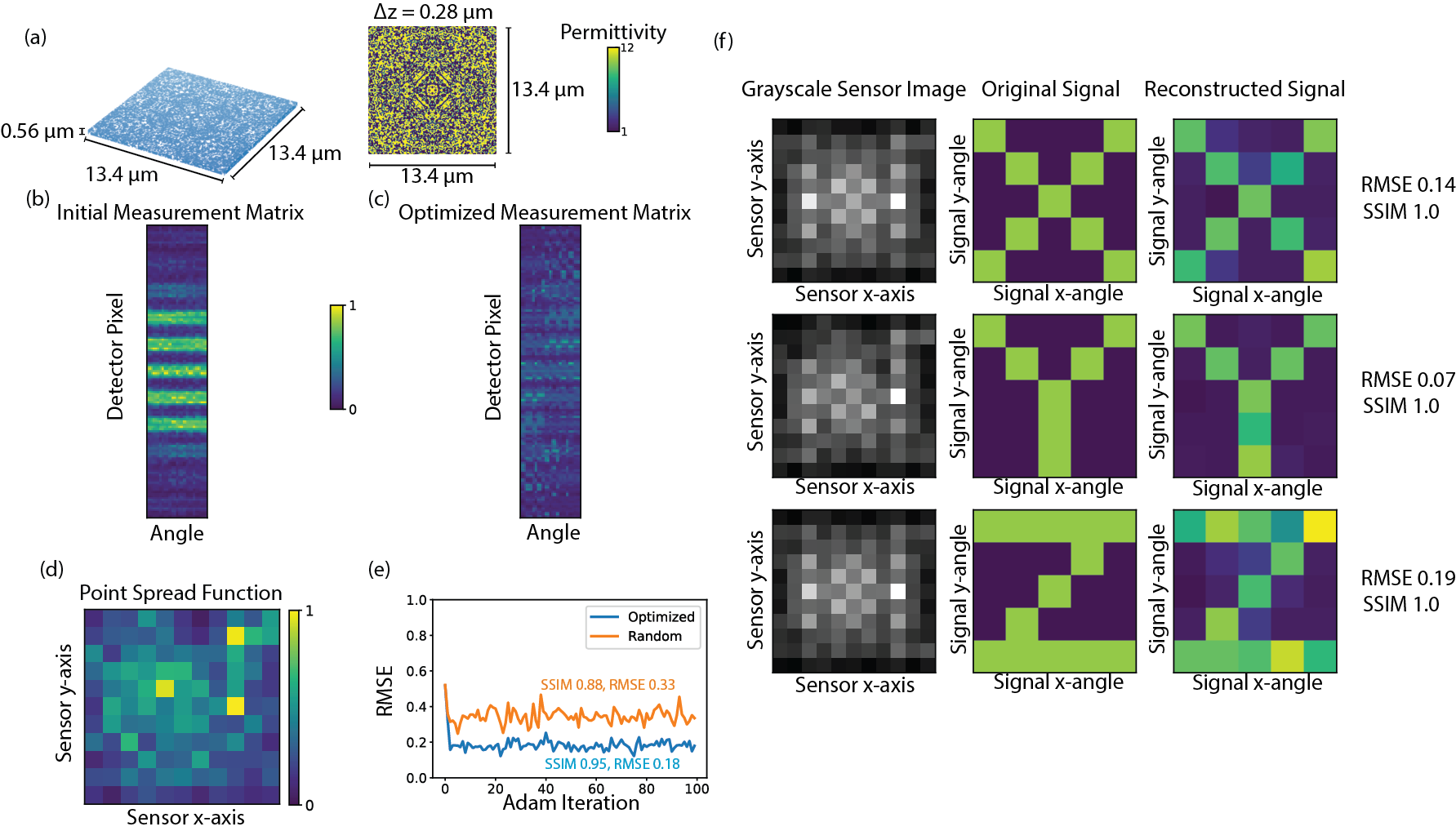"}
\caption{\textbf{3D Imager.} (a) Cross-sections of the optimized $13.4 \, \mathrm{\mu m }\times 13.4 \, \mathrm{\mu m }\times 0.56 \, \mathrm{\mu m }$ volumetric nanophotonic structure at $\Delta z =0.28 \, \mathrm{\mu m}$ from the surface of the nanophotonic structure, with 10648 design degrees of freedom per $1 \, \mathrm{\mu m}^3$. In the final optimized structure, $\varepsilon(\mathbf{r})$ takes on binary permittivities of only $1$ and $12$. (b, c) The measurement matrices $G$ (b) before and (c) after optimization. (d) Point spread function measured for the normal-incidence original signal. (e) Convergence of the reconstruction error during end-to-end optimization. (f) Side-by-side examples of grayscale sensor image, original signal, and reconstructed signal.}
\label{fig:4}
\end{figure}

Our last example reconstruction problem is a 3D extension of the result in Section \ref{2D_squeeze}. Here, like in the 2D space squeezing problem, the reconstruction problem is overdetermined. In this scenario, we set our ground-truth object to be of only one frequency and $5 \text{ x-angles}\times 5 \text{ y-angles}$, so the ground-truth vector $u$ is a length-$25$ vector. The angles here are spaced between $-0.02$ radians and $0.02$ radians from normal incidence in both the $x-$ and $y-$ dimensions. Each detector pixel is a square with side length $x_D = 2.24 \mathrm \, {\mu m}$. We set the sensor $d=11.2 \, \mathrm{\mu m}$ from the structure. We allow all $25$ components of the ground-truth to be activated in the optimization. We use an $11\times 11$ sensor, which makes the overall sensor have size $24.6 \, \mathrm{\mu m} \times 24.6\, \mathrm{\mu m}$, The nanophotonic design region is of size $13.4 \, \mathrm{\mu m} \times 13.4\, \mathrm{\mu m} \times 0.56\, \mathrm{\mu m}$ (\cref{fig:4}(a)), with 10648 design degrees of freedom per $1 \, \mathrm{\mu m}^3$. Here, we perform freeform optimization over voxels, as opposed to alternative of optimizing 2D patterns.

Before optimization, we initialize the nanophotonic as a random structure, which forms the measurement matrix shown in \cref{fig:4}(b). After optimization, the measurement matrix becomes the one shown in \cref{fig:4}(c) and the nanophotonic design becomes the structure shown in \cref{fig:4}(a). Qualitatively, the optimized measurement matrix is more sparse, and different angles are more focused on the sensor than in the initial measurement matrix, which serves to make the optimized matrix better-conditioned. Quantitatively, the condition number of the measurement matrix decreases from 902 to 221 over the optimization (Table 1). However, unlike in the 2D reconstruction problems, the transmission slightly decreases here, likely a result of the limited design space. We also show the point spread function for the optimized structure (\cref{fig:4}(e)). The intensity in the point spread function is localized in the top left corner. In the optimized measurement matrix (\cref{fig:4}(c)), the point spread function corresponds to the middle column, and the high intensity can be seen at the top of the row. Three example reconstructions are shown in \cref{fig:4}(f); qualitatively the reconstructed signals generally faithfully capture the main features of the original signals. Overall, the optimized system is benchmarked to have SSIM 0.95 and RMSE 0.18 at 1\% noise, a significant improvement over the system with a random structure (SSIM 0.88, RMSE 0.33).  

Lastly, this 3D structure achieves $\frac{x_D}{d} = 0.20$ and its compression ratio is $22.9$, beyond the realm of paraxial optics.

\begin{table}[!ht]
    \centering
    \begin{tabular}{|c|c|c|}
    \hline
         & Condition Number & Transmission \\\hline
        2D thin lens, sparse & 2.36$\times10^7$ & 0.90\\\hline 
        2D random structure, sparse & 2431 & 0.21 \\\hline
        \textbf{2D optimized structure, sparse} & 1514 & 0.27 \\\hline
        2D thin lens, space squeezing & $1.65 \times 10^{10}$ & 0.90 \\\hline
        2D random structure, space squeezing & 499 & 0.20 \\\hline
        \textbf{2D optimized structure, space squeezing} & 166 & 0.53 \\\hline 
        3D random structure, space squeezing & 902 & 0.39 \\\hline 
        \textbf{3D optimized structure, space squeezing} & 221 & 0.36 \\\hline
    \end{tabular}
    \\
    ~\\
    \caption{Condition numbers and transmissions over various designs.}
\end{table}

\begin{table}[!ht]
    \centering
    \resizebox{\textwidth}{!}{
    \begin{tabular}{|c|c|c|c|}
    \hline
         & Compression Ratio & Maximum Angle (radians) & Wavelength Range (nm) \\\hline
        Reshef et al., metamaterial spaceplate & 4.9 & 0.26 & 1550\\\hline 
        Reshef et al., uniaxial spaceplate & 1.12 & 0.61 & visible light (400--700) \\ \hline
        Guo et al., three-layer & 144 & 0.01& dependent on design constant (single wavelength) \\\hline
        Guo et al., single-layer hexagonal &11.2&0.11&dependent on design constant (single wavelength)\\\hline
        {\bf Our work, 2D (sparse and space squeezing)}&34.5&0.04&460--690\\\hline
 {\bf Our work, 3D (space squeezing)}&22.9&0.02&550

  \\\hline
    \end{tabular}}
    \\
    ~\\
    \caption{Comparison of our work with other space squeezing designs.}
\end{table}

\section{Discussion and Outlook}

\label{sec:discussion}

 Our central contribution is a flexible, noise-robust framework for transcending shift-invariance while imaging in the paraxial regime. By designing a volumetric nanophotonic structure with topology optimization for our optical element, we are no longer beholden to the shift-invariant paraxial approximation. Compared to traditional lenses or conventional metasurfaces (relying on the locally periodic approximation), we can keep the detector closer and the sensor resolution lower, which lets us keep the entire imaging system compact. We demonstrate that our method significantly outperforms both a thin lens and a random scattering structure in paraxial imaging with compression ratios of up to 34.5. In comparison, previous works have demonstrated compression factors of up to 4.9 \cite{reshef2021optic} and 144 \cite{guo2020squeeze}.
 
 We noted above in our examples (summarized in Table~1) that our optimization either preserves or improves transmission by the nanophotonic structure. For instance, the two-dimensional space squeezing optimization shows a 2.5x increase in transmission. This leads to a twofold benefit---the improved reconstruction accuracy shows that the system becomes more noise-robust, and the increase in transmission shows the system increases the signal-to-noise ratio.
 
 Future work may build on our framework by innovating on either the nanophotonic design or reconstruction algorithm. The choice of freeform nanophotonics opens our design to many more possibilities than prior works with locally-periodic metasurfaces, but there is a tradeoff in computational cost. For instance, to optimize a freeform nanophotonic structure in three dimensions (Section 3.3), we had to significantly reduce our problem size. With more computational power or more efficient design choices, the nanophotonic design could be scaled up to explore richer physics and higher-resolution imaging. For instance, this could be done with Flexcompute or by imposing an axisymmetric structure to reduce computational costs \cite{christiansen2020fullwave}. Innovations on the reconstruction algorithm may include replacing our elastic-net reconstruction with more general algorithms, such as neural networks. Furthermore, to bring our designs to physical reality, there are additional fabrication constraints to account for during topology optimization, such as minimum length scales and connectivity \cite{hammond2022high, hammond2021photonic}, which we do not yet account for in this present proof-of-concept work. 

Looking forward, we anticipate a growing demand for compact imaging. Our end-to-end framework coupled with freeform nanophotonics paves the way for the design of optical elements that can perform high-resolution imaging with limited volume. In particular, we present our method as a more general alternative to optics-only space squeezers. It is our hope that the application of end-to-end design to compact imaging will allow for smaller, higher-resolution cameras.

{\bf Funding.} G.A, W.F.Li, Z. Lin, C.R.C., M.S. and S.G.J. were supported in part by the U. S. Army Research Office through the Institute for Soldier Nanotechnologies under award number W911NF-18-2-0048.

\bibliographystyle{unsrt}  
\bibliography{references}
\end{document}